\documentclass[a4paper,11pt]{article}
\usepackage[latin1]{inputenc}
\usepackage{graphicx}
\usepackage{amsmath}

\newcommand{\vett}[1]{ {\mathbf{ #1}} }

\newcommand{\tdot}[1]{\hskip2pt\ddot{\null}\hskip 3.2pt \dot{\null}\kern -5pt {#1}}

\newcommand{\dif}{ \,\mathrm{d}}

\begin{document}

\title{Transition from order to chaos, and density limit, in   
magnetized plasmas}

\author{A. Carati\thanks{Universit\`a degli Studi di Milano, Milano,
    Italy. E-mail:   \texttt{andrea.carati@unimi.it}}\and
  M. Zuin\thanks{Consorzio RFX,   Associazione EURATOM-ENEA sulla
    Fusione, Padova,
    Italy. E-mail: \texttt{matteo.zuin@igi.cnr.it}}\and
  A. Maiocchi\footnotemark[1]\and M. Marino\footnotemark[1]\and
E. Martines\footnotemark[2]\and L. Galgani\footnotemark[1]}

\date{\today} 

\maketitle
 
\begin{abstract}
It is known that a plasma in a magnetic field, 
conceived microscopically as a system of point charges,  can exist in a 
magnetized state, and thus remain confined, inasmuch as it is in an 
ordered state of motion, with the charged particles  performing 
gyrational motions  transverse to the   field. 
Here we give an estimate of a threshold, beyond which   transverse
motion become chaotic, the electrons  being unable to perform 
even one gyration,
 so that a breakdown should occur, with complete loss of   confinement. The
estimate is obtained by the methods of perturbation theory,  
taking as perturbing force acting on each electron   that due to  the so--called
microfield, i.e.,   the  electric field produced by all the other
charges. We first obtain a general relation for the threshold, which
involves the fluctuations of the microfield. Then, taking for such 
fluctuations the fomula given by Iglesias, Lebowitz and MacGowan for
the model of 
a one component plasma with neutralizing background, we obtain a
definite formula for the threshold, which corresponds to a  density limit
increasing as the square of the imposed magnetic field. Such a
theoretical density limit is found to fit pretty 
well the empirical data for  collapses of fusion machines.
\end{abstract}



\section{Introduction}
The existence of a transition from order to chaos in Hamiltonian systems,
as a  generic phenomenon occurring when a perturbation is added to an
integrable system, is a well established fact. This fact  involves 
deep mathematical
features (see for example \cite{arnold}), and was made popular in the
scientific community through the striking pictures of  H\'enon and
Heiles \cite{henon} and the discovery, by Izrailev and
Chirikov \cite{ic}, that the ordered motions found by
 Fermi, Pasta and Ulam in their model \cite{fpu} become chaotic
above a certain stochasticity threshold (see Figs. 4.3 and 4.5 of the review  
\cite{giovanni},  or  \cite{fpuchaos}). 
 Transitions of this type  were met  also in the frame of 
plasma physics,  
in connection with the destruction  of magnetic
surfaces \cite{zasla,zasla2}, 
and also  with the chaoticity thus induced on
single particle motions \cite{neishtadt3}. 

On the other hand, in plasma physics  
a phenomenon  of  great  relevance exists that  is  yet unexplained,
and for which we propose here an explanation just in terms of a
   transition from order to chaos.
We refer  to the loss of  plasma confinement, a plasma collapse that
is met when the
plasma density is increased beyond a certain density limit
(see \cite{greenwald}, Fig. 3).

Let us recall  that confinement
amounts to keeping the charged particles away from the walls, 
and is actually achieved  by means of a suitable magnetic field,
the form of which depends  on the concrete machine (either just  a
field  imposed from outside, or a superposition of the imposed one
 with that due to a plasma current).
So, when the phenomenon of destruction of magnetic surfaces was
understood,  people thought that it might play a role in explaining  
the breakdown occurring at the density limit 
(see for example \cite{cargese}). The idea is that
the gyration centers should move, when the interactions are neglected,
along the field lines,  so that some peculiar topological features of
the latter,
particularly  in presence of  destructions  of the corresponding magnetic surfaces,
might be relevant. However, such considerations did not prove
sufficient to explain the quick collapses of plasmas.   

Here we propose a solution of a different character, completely  
unrelated to peculiaries of the field lines, up to the point of applying even 
in the  extremely idealized   case  in which the field is   uniform, so that the
 field 
lines  are just straight parallel lines, covering the whole space (and
the plasma is uniform too). 
The phenomenon
which we take into consideration is the existence of a magnetic pressure, 
which is essential in keeping the particles away from the walls. On
the other hand, 
it is well known that such a 
pressure exists inasmuch as the plasma is diamagnetic, which means, in 
microscocpic terms, that each  electron is equivalent to a magnetic 
moment, just  in virtue of 
its dynamical property of performing gyrational motions transverse to the field
lines. This is the kind of ordered motions we are referring to.

Now, such ordered motions induced by the external magnetic field,
persist indefinitely in the unperturbed case,  
when one neglects the  perturbation due to the so called microfield, i.e.,
the microscopic electric field acting on each charge  and due
to the Coulomb interactions with all the other ones (see the recent
review \cite{demura}). On the other hand the 
intensity  of such a perturbation clearly increases with the density, 
and so it seems
natural to expect that when the perturbation is large enough, i.e., at 
a large enough density, a transition to a state of chaotic motions should
occur,  
in which diamagnetism is lost, together with magnetic pressure. The proposal
advanced here is that such a kind of transition may explain the loss of 
confinement, at least in its gross features,  by providing a theoretical 
estimate of the density limit that should be  compatible, as far as  order of 
magnitude is concerned, with  the observed ones.

This is what we actually find out. Working with an extremely simple 
model, we predict a  stochasticity threshold, which  corresponds 
to a density limit  that  fits pretty well  
those  observed in collapses of several kinds of fusion machines.
 
A key point for  obtaining the stochasticity threshold
is the possibility of applying the 
methods of perturbation theory to macroscopic systems, as  a plasma is. 
Indeed, 
it is well known that the  classical estimates, which were brought to their
extreme range through the work of Nekhoroshev \cite{nek,bgg,galla},  
don't apply in the 
thermodynamic limit. In the present days, however, we can rely on the recent 
progress which was  obtained 
 when the classical scheme was 
extended as to hold in a probabilistic frame \cite{neishtadt2},
contenting oneself with controlling 
the  vast majority of  initial data (with respect to a given invariant
measure), rather than all of them.
Along such lines of research,   perturbation 
techniques were then extended  as to be  applicable in the thermodynamic 
limit \cite{andreajsp,aacmp,alberto}.

Using such techniques, we show that the zeroth order of
perturbation theory is already sufficient to produce a stochasticity 
threshold, which  turns out to fit pretty well the empirical data.
A characteristic feature of such new approaches, which  combine perturbation 
and statistical methods,  is that, in estimating the relaxation 
times or the stochasticity thresholds, the perturbing force enters through its 
fluctuations. In the present  case, what matters is the fluctuation
of the microfield, and we are in the fortunate situation that   a
theoretical formula for it is available (for the model of a one
component plasma with neutralizing background), which was  given   
already in the old work of Iglesias, 
Lebowitz and MacGowan \cite{iglesias2}. This finally  allows us to get the
theoretical formula for the density limit which is proposed here.

\section{Confinement and perturbation theory}
The idea that the loss of confinement in magnetized plasmas
corresponds to  a   transition
from order to chaos  is easily understood. 
Indeed an essential point
in guaranteeing confinement is the existence of a magnetic
pressure. Now,  in a macroscopic magnetohydrodynamic  description of 
the problem, the existence of a magnetic pressure is derived from the
constitutive equations of a plasma, i.e., from the assumption that the
plasma be diamagnetic.
However,  from a microscopic
point of view  in which the plasma is 
modeled as a system of discrete charges, such an assumption has to be
justified. In such a perspective,
 diamagnetism is a dynamical
property that can be present or absent, according  to the motions being
ordered or chaotic. In fact,  existence of  magnetization  corresponds
to the prevailing of gyrational  motions transverse to the
field, whereas in  the  state of statistical equilibrium (i.e., with
prevailing chaotic motions)  magnetization 
vanishes (see for example  \cite{vleck}). 
This breakdown  is thus  a global  characteristic   
feature of magnetized plasmas, irrespective of the particular
mechanism employed for obtaining confinement.

The conception that magnetization due to orbital motions can exist
only  in  a nonequilibrium state, 
characterized by  motions of ordered type,  was
apparently first proposed by Bohr (see \cite{bohr}, page 382). 
Now, Bohr  took  for granted that
the  relaxation time to equilibrium would be very short, as
\emph{``the collective motions of the electrons would disappear very
  rapidly''}. On the other hand, we are
well acquainted with the fact that the relaxation time 
from order to chaos can be very long, as occurs for example with
glasses and with the FPU model,  and was recently  pointed out also 
in connection with
orbital magnetization \cite{benfenati}. Thus, in order to establish up to which time
is  the magnetized state  conserved, one should
estimate a typical relaxation time after
which the system becomes  chaotic 
(see for example the \emph{``characteristic time of mixing''} defined in 
\cite{neishtadt}, sect. 5).

It is well known that this is a quite hard task.
However, estimates  of the relaxation time from below  are available
through perturbation theory, as we now recall.  Indeed, in general 
 such a theory
allows one to construct adiabatic invariants $I^{(n)}$ at any order
$n$,  providing for their  changes  $I_t^{(n)}-I^{(n)}$   (where $X_t$ denotes
  the time evolved at time $t$ of any  dynamical variable $X$)
 estimates which in their simplest form are  of the type
\begin{equation}\label{var_azione}
\left| I^{(n)}_t- I^{(n)}\right|\le n! \epsilon^{n+1}\, \mathcal I 
\frac t\tau\ ,
\end{equation}
where $\epsilon$ is the perturbation parameter, while  $\tau$ and
$\mathcal I$  are   a
characteristic time and  a characteristic value of $I$, of the
system. 
Now, imposing $\left| I^{(n)}_t- I^{(n)}\right|\le \mathcal I  \epsilon$, and
recalling $n! \simeq (n/e)^n$, formula (\ref{var_azione}) gives $t\le \tau
(e/n\epsilon)^n$ for all $n$, which, by taking the optimal value of
$n$, $n(\epsilon)\simeq 1/\epsilon$, gives  $t\le \tau \exp
(1/\epsilon)$.
Thus, a lower estimate
to the relaxation time is obtained, which is   exponentially long in
$1/\epsilon$ as long as $\epsilon<1$. It is thus clear that  the
condition $\epsilon=1$ provides a natural stochasticity threshold,
which should identify  the relevant transition, at least as
concerns the  order of magnitude of the characteristic parameters of
the problem.  
Indeed, for smaller $\epsilon$
the motions keep an ordered character for practically infinite times,
whereas for larger $\epsilon$  the ordered character is not even  guaranteed
up to the  microscopic time $\tau$.
As a matter of fact, the estimates for the changes of the adiabatic 
invariants are
in general a little more complicated than (\ref{var_azione}), and  the
lower estimates for the relaxation time  are found to increase as
stretched (rather than pure)  exponentials, but the
conclusion for  the stochasticity threshold to be drawn in a moment,
remains unaltered.

This classical scheme was implemented  in a probabilistic frame  in the paper
\cite{neishtadt2}.
Later the scheme was  shown to be applicable   also 
for systems of macroscopic sizes, i.e., in the so called thermodynamic 
limit \cite{andreajsp,aacmp,alberto}, which is an essential
point for our purposes. In such a probabilistic frame, one renounces
to control the changes of the adiabatic invariant  along all single 
trajectories, and  just controls mean properties with respect to a
given invariant measure in phase space. For example, one can look at the time
autocorrelation function $C_{I^{(n)}}(t)$ of the adiabatic invariant
at order $n$,  defined as usual by
$$
C_{I^{(n)}}(t)=\langle I_t^{(n)}I^{(n)}\rangle-\langle I^{(n)}\rangle^2\ ,
$$
where $\langle \cdot \rangle$ denotes  mean with respect to the given 
measure. In terms of the time autocorrelation function,
the analogue of the classical estimate (\ref{var_azione}) then takes the form
$$
\frac{C_{I^{(n)}}(t)}{\sigma^2_{I^{(n)}}}\ge 1-\frac 12 \ n! \epsilon^{n+1}
\left(\frac t\tau\right)^2 \ ,
$$
where $\sigma^2_X=\langle X^2\rangle-\langle X\rangle^2$ is the
variance of $X$. The latter  provides  a natural dimensional constant for
the autocorrelation, since one has  $C_X(0)=\sigma_X^2$.  
By optimization with respect to $n$, the time after which the adiabatic
invariant may lose correlation is still found to be exponentially long 
in $1/\epsilon$,
provided one has $\epsilon<1$. So,   $\epsilon=1$ again turns   out  to be
 the perturbation estimate of the stochasticity threshold.

Our main task is thus to estimate the stochasticity threshold 
for a magnetized plasma, in the probabilistic frame just sketched. In
section~\ref{nuova} we describe the model we study, define the
 dynamical variable of interest (the  component of the angular momentum of each
electron along the field) and give the lowest order  estimate for  its time
autocorrelation function. This leads to a natural  conjecture
for identifying the
perturbation parameter $\epsilon$, which then gives the stochasticity 
threshold by the condition  $\epsilon=1$. In section~\ref{due} we give a 
general formula for the threshold
in terms of temperature and of the fluctuations of the microfield. 
Using the available  analytical estimate of such
fluctuations for the  model of a one component plasma with a neutralizing 
background \cite{iglesias2}, a 
definite formula  for the   theoretical density limit is then obtained. In
section~\ref{tre} the theoretical density limit is compared to  the
empirical data for collapses in fusion machines. Some comments are
finally  added in the conclusions.

\section{Definition of the model. Conjecture on the stochasticity
  threshold}\label{nuova} 
The   model chosen for the magnetized  plasma  is the simplest
one we could  conceive in order to  check  the main idea of the present paper, 
namely, that the relevant  feature concerning loss of
confinement, regardless of the particular mechanism involved in
each machine, is the occurring of a sharp  transition from order to chaos as the
perturbation due to the microfield is increased  beyond a 
threshold. So, first of all, for what concerns the magnetic field
$\vett B$,  we consider the extremely idealized case in which 
it is uniform, say   $\vett B=B\vett e_z$, 
where $\vett e_z$ is the unit vector along the $z$  axis. 
Concerning the plasma itself, the  key
point is  that it  should be 
conceived  as a dynamical
system of point charges, and not as a continuum. Thus, any charge will
be subiect, in addition to the Lorentz confining force due to $\vett
B$, also to the
force of the microfield  $\vett E$, defined  as the vector sum of the Coulomb
fields created by all the other charges. Mutual magnetic forces and
retardation effects are neglected.

So we have a system of
several kinds of charges, and  the Newton equation for the $j$th
charge  (in the nonrelativistic approximation) is then
\begin{equation}\label{uno}
m_j\ddot{\vett x}_j= e_j\vett v_j\wedge \vett B +e_j \vett E_j
\end{equation}
where $m_j$ and $e_j$ are the mass and the
charge of the particle, $\vett x_j$  and $\vett v_j=\dot {\vett x}_j $ 
its   position vector and velocity, and $\vett E_j=\vett E(\vett x_j)$ 
the microfield (evaluated at $\vett x_j$,  which depends on the positions   
of  all the charges. Finally, in order that the dynamical
system  be defined within  the standard approach of ergodic theory, 
we consider as given also  an invariant
measure, a few minimal properties of which will be mentioned later.

If the microfield is
neglected, the transverse motion of each particle is a   uniform
gyration about a field line  with its  characteristic 
cyclotron frequebcy $\omega_{cj}=|e_j|\, B/m_j$. So the system is
integrable, the $z$ component of the angular momentum  of each
particle being a constant of motion. The microfield,
acting as a perturbation, makes the system no more
integrable.

For what concerns the adiabatic invariant to be investigated, in
principle we should look at the magnetization of the system, to
which each charge contributes through  the $z$ component of its angular
momentum.
However, it is well known that only the electrons are
relevant, the ions contribution to magnetization  being  negligible.
So we will consider the contribution to magnetization due to
any single electron,  i.e., the $z$ component of its angular momentum.
Since now on,   the index $j$ referring to a chosen electron
will be left   understood. Thus,
as zeroth order approximation for the adiabatic invariant
we take the quantity
\begin{equation}\label{momento}
 L = \frac{m^2 }{e B}\, v^2_{\bot}
\end{equation}
($\vett v_\bot$ denoting transverse velocity of the chosen electron),
which is  
proportional to the transverse kinetic energy of the electron. 
One immediately checks (see
page 16 of the book of  Alfv\'en  \cite{alfven}, or  
any  plasma physics textbook) that
 $L$ is the $z$ component of the angular momentum  of the chosen
electron, referred to
its instantaneous  gyration center (or guiding center), the latter
being calculated in
the approximation in which the perturbing force is neglected

We also add here the formula for the time derivative
$\dot L$ of $L$, as  we will need it in a moment.
As  $L$ is a multiple of $\vett v_\bot \cdot \vett v_\bot$, 
$\dot L$ is immediatley obtained through dot multiplication of   
Newton's equation (\ref{uno}) by
$2\vett v_\bot$, which gives 
\begin{equation}\label{momento2}
\dot L= \frac {2m}{ B}\, \vett v_{\bot}\cdot \vett E_{\bot}\ .
\end{equation}

We come now to the main point: to  find the dimensionless perturbation parameter
$\epsilon$ which determines the stochasticity threshold
corresponding to the destruction of the chosen adiabatic invariant
(and of all the adiabatic invariants corresponding to each electron).
This would
require  performing the corresponding perturbation estimates  at all
orders, which at the moment we are unable to do. What we can
easily do is to perform the zeroth order estimate for the
time autocorrelation function of $L$, which turns out to be 
\begin{equation}\label{bbb2}
\frac{C_L(t)}{\sigma^2_L}\ge 1-
\frac12\frac{\sigma^2_{\dot
    L}}{\omega^2_c\sigma^2_L}\left(\omega_c t\right)^2\ .
\end{equation}
Notice that as  characteristic microscopic time $\tau$ of the
unperturbed electron's motion we have  naturally taken  
$1/\omega_c$.

The proof of (\ref{bbb2}) is rather simple. One starts from the elementary  identity
$$
C_L(t)=\sigma^2_L-\frac 12 \langle \left(L_t-L\right)^2\rangle\ ,
$$
and uses  the inequality $\langle \left(L_t-L\right)^2\rangle\le
\langle\dot L^2\rangle t^2$, which is just a function theoretic analogue of the
Lagrange finite increment formula  of elementary calculus, and
basically follows from 
unitarity of the time evolution of the dynamical variables (see  
\cite{andreajsp},  Theorem~1, and \cite{aacmp}, Section~7). This
already gives inequality (\ref{bbb2}), with
 $\langle \dot L^2\rangle$ in place of $\sigma^2_{\dot
  L}$. Relation (\ref{bbb2})  then immediatley follows by noting that, 
due to the time--invariance of the measure, for
any dynamical variable $X$  one has $\langle \dot X\rangle=0$, 
so that $\langle \dot X^2\rangle=\sigma^2_{\dot X}$.

As explained in the Introduction, from (\ref{bbb2}) we are led to  
conjecture that the relevant dimensionless parameter of the problem is 
\begin{equation}\label{epsilon}
\epsilon=\frac{\sigma_{\dot     L}}{\omega_c\sigma_L}\ ,
\end{equation}
($\sigma_X=\sqrt{\sigma_X^2}$ denoting standard deviation), and this
leads to a stochasticity threshold  given   by
\begin{equation}\label{soglia1}
\frac{\sigma_{\dot     L}}{\omega_c\sigma_L}= 1\ .
\end{equation}

We add now a comment  of a general
character, which  concerns the way in which a relaxation time for $L$
(or analogously for  any variable $X$) turns out to be identified in
the present approach, which combines  perturbation  and
statistical mechanics methods.   From (\ref{bbb2}) one sees that the
relaxation time $t_L^{rel}$ of $L$ is given by
$t_L^{rel}=\sigma_L/\sigma_{\dot L}$, a formula which involves
standard deviations. Now, compare such a formula with the  one generally  met 
in textbooks, i.e., $t_L^{rel}=L/\dot L$, where it should be   understood   
that ``typical values'' are to be taken for the
numerator and the denominator. But this requires a great ingenuity
from the part of the reader,
especially when variables are involved which have vanishing mean. So
one may say that the identification  of the  relaxation time  provided by
perturbation theory in a probabilistic frame, namely, 
$t_L^{rel}=\sigma_L/\sigma_{\dot L}$, appears to be some
definite quantitative  implementation of the intuitive idea underlying the
familiar informal definition, and amounts to the prescription that the
informal qualification ``typical values'' should be understood in the
sense of ``standard deviations''. 

We add now  a final remark in which  the previous
comment is used to read in a quite transparent way the condition (\ref{soglia1})
which defines  the stochasticity  threshold. Indeed, through formula
(\ref{soglia3})
of the next section it will be seen that the condition for the
threshold can be expressed in the form
$(\sigma_{E_\bot}/B\sigma_v)=1$. Thus, just in virtue of the previous
comment relating  standard deviations and  typical values, one sees that the 
 threshold occurs when the typical value
of the perturbing force due to the  microfield  equals
the typical value of the Lorentz force which characterizes the unperturbed motions.
So the condition   $\epsilon =1$, which we have assumend
 as a definition of the threshold within  a rather abstract point of
 vue, is just   what    
one would   immediately guess,
as the naivest implementation of the idea that a threshold occurs when
the perturbing force equals the unperturbed one.

\begin{figure*}[ht]
  \begin{center}
    \includegraphics[width=\textwidth]{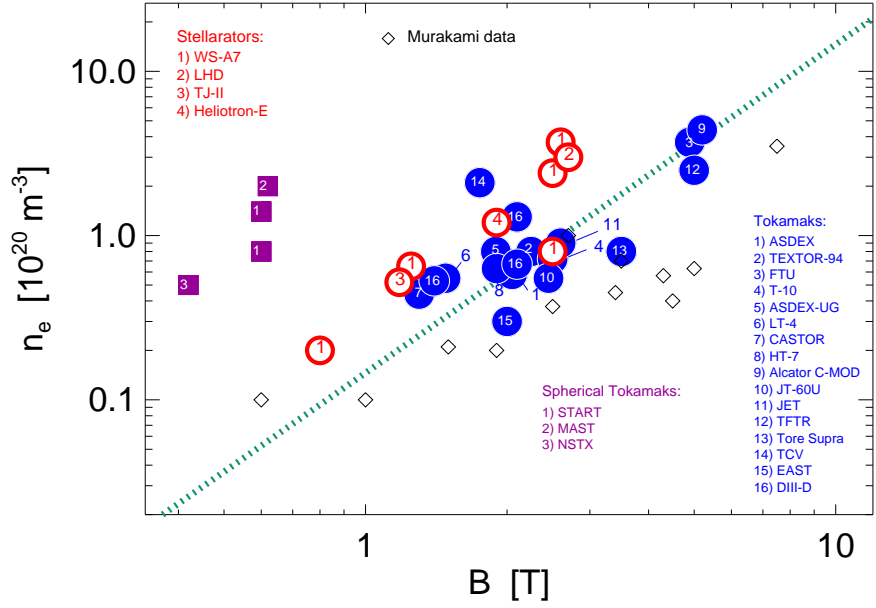}
  \end{center}
  \caption{\label{fig:1} Density limit values \emph{vs}  $B$ for
    various machines:  conventional tokamaks, for which recent data are
    shown (see references
    \cite{Stabler,DeVries,Frigione,Merezhkin,
      Merthens,Howard,Dyabilin,Asif, Bombard,Takenaga,Saibene,Bell,
      Commaux,Weisen,Liu, Petrie}) along with the original ones of Murakami (see
    \cite{murakami}),  stellarator
    machines \cite{Giannone,Mijazawa,Tabares, Sudo}, and spherical
    tokamaks \cite{Sykes_START,Martin_MAST, Kaye_NSTX}. Dotted line is
    the theoretical   density limit     (\ref{sei}). }  
\end{figure*}

\section{The stochasticity threshold in terms of macroscopic
  parameters. The theoretical density limit}\label{due}

Our aim is now to express    the stochasticity threshold 
(\ref{soglia1})
in terms of the macroscopic
parameters $T$, $n$, $B$, temperature, electron number density  and field
strength. Recalling the expressions (\ref{momento}), (\ref{momento2})
of $L$ and $\dot L$,  and the definition
$\omega_c=|e|\, B/m$, the threshold  (\ref{soglia1}) takes the form 
\begin{equation}\label{soglia0}
\frac 2B\,\frac{\sigma_{\vett v_\bot\cdot \vett E_\bot}}{ \sigma_{v^2_\bot}}=1\ .
\end{equation}

It is clear that the  standard deviations  appearing in (\ref{soglia0}) 
depend on the model of plasma adopted, which determines the
microfield, as well as  on the chosen invariant
measure. The choice of the invariant measure is a quite delicate
problem, particularly in a nonequilibriun situation as the one we are
discussing here. A general introduction may be found in the book 
\cite{giovanni2}.
 For example, it is obvious that  $\sigma_{\vett   v_\bot}$ and
  $\sigma_{v^2_\bot}$  should be
expressed in terms of temperature, albeit with coefficients  which
depend  on the assumptions made for the velocity
distribution. Analogously, the statistical properties of the microfield
may be different for a system  composed by electrons plus  a  neutralizing
background, rather than for  a system of electrons and ions.

Quite natural assumptions on the measure are: i) 
that velocities and positions are independent variables;  
ii) that the distribution of the transverse velocities is  Maxwellian 
at a temperature $T$; iii) that the distribution of positions is isotropic.
Under  these  natural assumptions 
the equation (\ref{soglia0}) for the threshold is seen to take the form
\begin{equation}\label{soglia3}
 \sqrt{\frac 23}\,\, \frac {\sigma_{\vett E}}{B\,  \sqrt{k_BT/m}}=1
\end{equation}

Indeed, from i) and ii) one gets
$$
\sigma^2_{\vett v_\bot\cdot \vett E_\bot}= \frac 12\, 
\sigma^2_{\vett  v_\bot}\, \sigma^2_{\vett  E_\bot} \ ,
$$ 
the variance of a vector $\vett F$ being defined   by $\sigma^2_{\vett
  F}=\sigma^2_{F_x}+\sigma^2_{F_y}+\sigma^2_{F_z}$.
One also gets
$$
 \frac 12\, 
\sigma^2_{\vett  v_\bot}=
\frac{k_BT}m
$$
 ($k_B$ being  the Boltzmann constant) and furthermore, as one easily checks, 
$$
\sigma^2_{v_\bot^2}=
4\left(\frac{k_BT}m\right)^2\ .
$$
Finally, from iii) one gets
${\sigma^2_{\vett E_\bot}=(2/3)\, \sigma^2_{\vett E}}$

The form  (\ref{soglia3}) of the equation for   the thresold
already constitutes in  our opinion a significant result.
Indeed, the fluctuation $\sigma^2_{\vett E}$ of the microfield
 should in principle be itself  a measurable
quantity, which depends on the macroscopic state of the plasma, namely,
electron number density $n$ and the temperatures of the several
constituents. So the previous relation provides in principle the
density limit as a function of the macroscopic state of the plasma.

However,  we were unable to  find in the  literature  sufficient
experimental information  on  the fluctuation $\sigma^2_{\vett E}$ of 
the microfield. So, in order to have a definite
theoretical formula to be compared with the experimental data, we 
limit ourselves to the consideration of a particular model for which 
an estimate of $\sigma^2_{\vett E}$ is available. In fact a formula
for  $\sigma^2_{\vett  E}$ was given by Iglesias, Lebowitz and 
MacGowan \cite{iglesias2}
for the model of a one component plasma with neutralizing background, 
at equilibrium  
with respect to the Gibbs  distribution, for which they  found  
\begin{equation}\label{lebo}
\sigma^2_{\vett E}= \frac{ n\,  k_B T}{\varepsilon_0}\ ,
\end{equation}
where  $\varepsilon_0$ is the vacuum dielectric constant and $n$  the
electron number density (see \cite{iglesias2}, formula
(2.5), substituting $n$ for  $\rho$ and   $1/\varepsilon_0$ for
$4\pi$).  

So, for a one component plasma with neutralizing background at Gibbs
equilibrium  at temperature $T$, the   stochasticity  threshold
(\ref{soglia3})  takes the form 
\begin{equation}\label{sei}
n= \frac 32\, \frac{\varepsilon_0}m \,B^2\ ,
\end{equation}
in which temperature disappeared, so that the threshold only involves
density and field strength. 
Notice however that this might not be true with a more realistic model
of a plasma, in which the temperature appearing in (\ref{lebo}), which
refers to the plasma as a whole, may be different from the electron
transverse temperature which enters the previous formulas.

Formula (\ref{sei}) for the limit density (holding for a one component
plasma with neutralizing background, at Gibbs
equilibrium  at temperature $T$) is the type of result we were looking 
for, inasmuch as it  provides a definite
theoretical formula for the density limit that can be compared to the
 available empirical data for collapses in fusion machines, as will be
 done in the next section.

\section{Comparison with the
  empirical data for plasma collapses in fusion machines}\label{tre}

We now  check whether the transition from order to chaos 
discussed here has anything to do with the empirical 
data for collapses in fusion machines.  
We recall that a proportionality of the  density limit to the
square of the magnetic field in tokamaks  was  
suggested by Granetz  \cite{granetz} on the
basis of empirical data, but apparently was not confirmed by later
observations  \cite{Petrie,greenwald}. 
It is well known that, while at first a proportionality 
to the magnetic field (through $B/R$, where $R$ is the major radius of
the torus)  
had been proposed  on an empirical basis for tokamaks  by
Murakami  \cite{murakami}, in the plasma physics community the common 
opinion is rather that
the  density limit for tokamaks should be proportional to the
Greenwald parameter $I_p/r_a^2$, where $I_p$ is the plasma current and
$r_a$ the minor radius of the torus (see \cite{greenwald}). 

We do not enter here a
discussion of this point, and only content ourselves with plotting in 
Figure \ref{fig:1}
a collection of available data of  the  density limit  for several
fusion machines  versus
their operating magnetic field $B$ in log--log scale, comparing the data to the
theoretical formula (\ref{sei}).
The first  thing that comes out from the figure is that the order
of magnitude of the  theoretical threshold is  correct, and this without
having introduced  any phenomenological parameter. There is no
adjustable parameter in the theory, and no fittind at all.
One is thus tempted to say that the essence of the phenomenon has
perhaps been captured, 
especially in consideration of the extreme
simplicity of  the model, with respect to  the  
variety  of  machines and of  operational conditions  to which the experimental
data refer.

Entering now in some more details, one sees that the theoretical 
law appears to correspond not so
badly to the data  for the high field machines (tokamak and
stellarators), whereas a sensible discrepancy is met for the low
field machines (spherical tokamaks), for which  the 
experimental data are larger by even an order of magnitude.
Perhaps this discrepancy might be attributed  to the fact that
 we are discussing here a model
describing  an isolated, non sustained, system (i.e., with no input heating
power), whereas  
the low field machines considered
in the figure
are just the ones characterized, in general, by lower
confinement time and thus by larger sustainment. Indeed,
(see the empirical Sudo limit for
stellarators \cite{Sudo})  larger densities are expected to be  accessible as   the 
 input power is increased (although this is not so clear for tokamaks 
 \cite{greenwald}). This is illustrated, in the figure, by the
 three points reported for  the same  device (the stellarator 
WS-A7  \cite{Giannone})  at  
 essentially the same applied field, which however  correspond to  three 
different (increasing)  input heatings.  

\section{Conclusions}
In view of the lack of any first principles rationale for  the
existence of a density limit in fusion machines, the comparison
between theory and experiments exhibited in Figure \ref{fig:1}  
appears  encouraging.  Particularly so, if one considers the extreme
simplicity   of the model (uniform plasma in a uniform field), with respect to  the
the  variety  of  machines and of  operational conditions  to which
the experimental data refer. The essence of the phenomenon seems to 
have been captured.

So, one might  consider as plausible 
the main proposal advanced in the present paper, namely,   that
the density limit  characterizing the empirical collapses of  fusion machines 
corresponds to a  transition from order to chaos in the following sense. 
At low densities, 
ordered motions due to the  imposed  magnetic field  prevail, with 
the electrons performing transverse  gyrational motions, and thus with 
a magnetic pressure. Then,  as density is
 increased, the
perturbations caused by the fluctuations of the microfield
(which increase as the  density) introduce a 
chaotization, until a
stochasticity limit (and so a  density limit) is attained, 
beyond which ordered motions are lost, together  with magnetic 
pressure and confinement. 

A   key feature of the present
approach, with respect to  treatments 
involving the continuum approximation, such as magnetohydrodynamics, 
is that we are dealing here with the
plasma as a discrete system of 
charges.  Indeed in our treatment an essential   role is played by  the 
microfield acting on
a single electron, and so  it is not clear how 
the instability found here could find place
within  the continuum approximation,  or  any
other approximation involving high--frequency cutoffs. 
For an analogous role of discreteness of matter
in cosmology, see \cite{ccg} and \cite{galassie}.

Actually, even in plasma physics theory there exists a huge
literature  in which the discrete nature of matter is taken 
into account, following the  approach of kinetic theory 
(see for example \cite{ichimaru}). A comparison with the
 results obtained here within the approach of  dynamical systems theory 
would thus be in order. We hope to come back to 
this problem in the future.

A further remark is that the existence of a  density limit
proportional to the square of the magnetic field is well known in the
frame of nonneutral plasmas (see \cite{davidson}), under the
name of Brillouin limit. The physical
context is however  rather different, because  the density limit  in
the latter  case  refers to  the existence of a
particular motion, in which   the plasma, dealt with as a  continuum, 
performs a rigid rotation 
about the $z$ axis. Actually, it is clear that a magnetization threshold in 
the sense discussed here should exist  for nonneutral
plasmas too. The only problem is that we are unaware of any estimate of
the standard deviation of the microfield in such a case. We hope to
come back to this problem in the future.

We finally  point out that the proportionality of the density limit
to the square of the magnetic field predicted  by the theoretical law
(\ref{sei}), if confirmed,
might have relevant implications for future tokamaks.

\vskip .2truecm
\textbf{Acknowledgements.}
This paper is dedicated to Giovanni Gallavotti, on the occasion of his
seventieth birthday.
We also thank  N. Vianello for fruitful discussions.


\begin{thebibliography}{99}
\bibitem{arnold} V.I.  Arnold (Ed) Dynamical Systems III, Encycolpedia of
  Mathematical Sciences, Volume 3, Springer--Verlag (Berlin, 1988).
\bibitem{henon} M. H\'enon, C. Heiles, \textsl{Astron. J.}
  \textbf{69}, 63 (1964).
\bibitem{ic} F.M. Izrailev, B.V. Chirikov, \textsl{Sov. Phys. Dokl.}
  \textbf{11}, 30 (1966).
\bibitem{fpu} E. Fermi, J. Pasta, S. Ulam, in \textsl{E. Fermi:
Note e Memorie (Collected Papers)}, Vol. II, N. 266: 977 (Accademia
Nazionale dei Lincei, Roma, and The University of Chicago Press, Chicago
1965).
\bibitem{giovanni}  G. Benettin, A. Carati, L. Galgani, A, Giorgilli,
  in G. Gallavotti ed., \textsl{The Fermi-Pasta-Ulam
  Problem: A Status   Report}, Springer Verlag (Berlin,2007).
\bibitem{fpuchaos} A. Carati, L. Galgani, A. Giorgilli, \textsl{Chaos}
  \textbf{15}, 015105 (2005). 
\bibitem{zasla} M.N. Rosenbluth, R.Z. Sagdeev, J.B. Taylor,
  G.M. Zaslavsky, \textsl{Nucl. Fusion} \textbf{6}, 297 (1966).
\bibitem{zasla2} N.N. Filonenko, R.Z. Sagdeev, G.M. Zaslavsky,
  \textsl{Nucl. Fusion}  \textbf{7}, 253 (1967).
\bibitem{neishtadt3} A. Vasiliev, A. Neishtadt, A. Artemyev,
  \textsl{Phys. Lett. A} \textbf{375}, 3075 (2011).
\bibitem{greenwald} M. Greenwald, \textsl{Plasma Phys.
  Contr. Fusion} \textbf{44}, R27 (2002).
\bibitem{cargese} G. Laval, D. Gresillon eds., \textsl{Intrinsic
  stochasticity in plasmas}, Les Editions de Physique, Courtabeuf
  (Orsay, 1980). 
\bibitem{demura} A.V. Demura, \textsl{Intl. J. Spectroscopy}
  \textbf{2010}, 671073, (2010).
\bibitem{nek} N.N. Nekhoroshev, \textsl{Russ. Math. Surveys}
  \textbf{32} N.6, 1 (1977).
\bibitem{bgg} G. Benettin, L. Galgani, A. Giorgilli,
  \textsl{Cel. Mech.} \textbf{37}, 1 (1985).
\bibitem{galla} G. Gallavotti, \textsl{Commun. Math. Phys.}
  \textbf{164}, 145 (1994).
\bibitem{neishtadt2} A.I. Neishtadt, \textsl{Sov. Phys. Dokl.}
  \textbf{21}, 80 (1976).
\bibitem{andreajsp} A. Carati, \textsl{J. Stat. Phys.} \textbf{128}, 1057 (2007).
\bibitem{aacmp} A. Carati, A. Maiocchi, \textsl{Commun. Math. Phys.}
  (2011), in press.
\bibitem{alberto} A. Maiocchi, A. Carati, \textsl{Commun. Math. Phys.}
  \textbf{297}, 427  (2010).
\bibitem{iglesias2} C.A. Iglesias, J.L. Lebowitz, D. MacGowan,
  \textsl{Phys. Rev. A} \textbf{28}, 1667 (1983).
\bibitem{vleck} J. H. Van Vleck, \textsl{The Theory of Electric and
  Magnetic Susceptibilities}, Oxford University Press (London, 1932).
\bibitem{bohr} N. Bohr, \textsl{Collected Works, Volume I: Early Works
  (1905--1911)},  L. Rosenfeld and J. Rud Nielsen  eds., North-Holland 
(Amsterdam, 1972).
\bibitem{benfenati} A. Carati, F. Benfenati, L. Galgani,
  \textsl{Chaos} \textbf{21},  023134 (2011). 
\bibitem{neishtadt} D.I. Vainchtein, J. B\"uchner, A.I. Neishtadt,
  L.M. Zeleny, \textsl{Nonlinear Processes in Geophysics},
  \textbf{12}, 101 (2005).   
\bibitem{alfven} H. Alfv\'en  \textsl{Cosmical electrodynamics},
  Oxford U.P. (Oxford, 1950).
\bibitem{giovanni2} G. Gallavotti, \textsl{Statistical mechanics. A
  short treatise},  Springer Verlag (Berlin, 1999).
\bibitem{granetz} R.S. Granetz, \textsl{Phys. Rev. Lett.} \textbf{49},
  658 (1982).
\bibitem{murakami} M. Murakami, J.D. Callen,  L.A. Berry,
  \textsl{Nucl. Fusion} \textbf{16}, 347 (1976).
\bibitem{Stabler} \"A. Stabler,  et al., \textsl{Nucl. Fusion}
    \textbf{32},  1557 (1992).
\bibitem{DeVries} P.C. de Vries, J. Rapp, F.C. Sch\"{u}ller,
  M.Z. Tokar, \textsl{Phys. Rev. Lett.} \textbf{80},  3519 (1998).
\bibitem{Frigione} D. Frigione, et al., \textsl{Nucl. Fusion} 
\textbf{36}, 1489 (1999).
\bibitem{Merezhkin} V.G. Merezhkin, \textsl{33rd EPS Conference on 
Plasma Phys.}, Rome 19-23 June 2006 ECA Vol. 30I, P-4.085 (2006).
\bibitem{Merthens} V. Merthens, et al.,  
\textsl{Nucl. Fusion} \textbf{37}, 1607 (1997).
\bibitem{Howard}J. Howard, M.  Person,   
\textsl{Nucl. Fusion} \textbf{32}, 361 (1992).
\bibitem{Dyabilin} K.S. Dyabilin, et al.,  
\textsl{Czech. J. Phys.} \textbf{37}, 713 (1987).
\bibitem{Asif} M. Asif, et al., \textsl{Phys. Lett. A}  
\textbf{336}, 61 (2005).
\bibitem{Bombard} B. LaBombard, et al.,  
\textsl{Phys. Plasmas} \textbf{8}, 2107 (2001).
\bibitem{Takenaga}H. Takenaga, et al.,  
\textsl{Nucl. Fusion} \textbf{45}, 1618 (2005).
\bibitem{Saibene} G. Saibene, et al. 
\textsl{Nucl. Fusion} \textbf{39}, 1133 (1999).
\bibitem{Bell} M.G. Bell, et al., \textsl{Nucl. Fusion} \textbf{32},
  1585 (1992).
\bibitem{Commaux} N. Commaux, et al., \textsl{33rd EPS Conference
  on Plasma Phys.}, Rome, 19 - 23 June 2006 ECA, Vol.30I, P-5.105 (2006).
\bibitem{Weisen} H. Weisen, et al. \textsl{Plasma Phys. Control. 
Fusion} \textbf{38}, 1137 (1996). 
\bibitem{Liu} H.Q. Liu, et al., \textsl{Plasma Phys. Control. 
Fusion} \textbf{49}, 995 (2007). 
\bibitem{Petrie} T.W. Petrie, A.G. Kellman, M.Ali Mahdavi, 
\textsl{Nucl. Fusion} \textbf{33}, 929 (1993).
\bibitem{Giannone} L. Giannone  et al.,  \textsl{Plasma
   Phys. Control. Fusion} \textbf{42}, 603 (2000).
\bibitem{Mijazawa} J. Mijazawa, et al.,
  \textsl{Nucl. Fusion} \textbf{48}, 015003 (2008).
\bibitem{Tabares} F.L. Tabar\'es, et al., \textsl{Plasma
  Phys. Control. Fusion} \textbf{50}, 124051 (2008).
\bibitem{Sudo} S. Sudo . et al., \textsl{Nucl. Fusion} \textbf{30}, 11 (1990).
\bibitem{Sykes_START} A. Sykes  \textsl{ Nucl. Fusion} \textbf{39},
  1271 (1999).
\bibitem{Martin_MAST} R. Martin et al., \textsl{27th EPS Conference on
  Contr. Fusion and Plasma Phys.}, Budapest, 12-16 June 2000 ECA,
  Vol. 24B, 221 (2000).
\bibitem{Kaye_NSTX}  S.M. Kaye  et al.,  \textsl{Phys. Plasmas}
  \textbf{8}, 1977 (2001).
\bibitem{ccg} A. Carati, S. Cacciatori, L. Galgani,
  \textsl{Europh. Letters} \textbf{83}, 59002 (2008).
\bibitem{galassie} A. Carati, \textsl{Gravitational effects of
  the faraway matter on the rotation curves of spiral galaxies}, 
 arXiv: 1111.5793. 
\bibitem{ichimaru} S. Ichimaru, \textsl{Statistical plasma physics},
  Addison Wesley, (Redwood City, 1992).
\bibitem{davidson} R.C. Davidson, \textsl{Physics of nonneutral
  plasmas}, Addison Wesley (Redwood City, 1990).
\end{thebibliography}
\end{document}